\documentclass[prd,12pt,tightenlines,nofootinbib]{revtex4}

\usepackage{amsmath,amssymb}

\begin{document}

\def\CA{{\cal A}}
\def\CB{{\cal B}}
\def\CC{{\cal C}}
\def\CD{{\cal D}}
\def\CE{{\cal E}}
\def\CF{{\cal F}}
\def\CG{{\cal G}}
\def\CH{{\cal H}}
\def\CI{{\cal I}}
\def\CJ{{\cal J}}
\def\CK{{\cal K}}
\def\CL{{\cal L}}
\def\CM{{\cal M}}
\def\CN{{\cal N}}
\def\CO{{\cal O}}
\def\CP{{\cal P}}
\def\CQ{{\cal Q}}
\def\CR{{\cal R}}
\def\CS{{\cal S}}
\def\CT{{\cal T}}
\def\CU{{\cal U}}
\def\CV{{\cal V}}
\def\CW{{\cal W}}
\def\CX{{\cal X}}
\def\CY{{\cal Y}}
\def\CZ{{\cal Z}}

\newcommand{\todo}[1]{{\em \small {#1}}\marginpar{$\Longleftarrow$}}
\newcommand{\labell}[1]{\label{#1}\qquad_{#1}} 
\newcommand{\bbibitem}[1]{\bibitem{#1}\marginpar{#1}}
\newcommand{\llabel}[1]{\label{#1}\marginpar{#1}}

\newcommand{\sphere}[0]{{\rm S}^3}
\newcommand{\su}[0]{{\rm SU(2)}}
\newcommand{\so}[0]{{\rm SO(4)}}
\newcommand{\bK}[0]{{\bf K}}
\newcommand{\bL}[0]{{\bf L}}
\newcommand{\bR}[0]{{\bf R}}
\newcommand{\tK}[0]{\tilde{K}}
\newcommand{\tL}[0]{\bar{L}}
\newcommand{\tR}[0]{\tilde{R}}

\newcommand{\ack}[1]{[{\bf Ack!: {#1}}]}

\newcommand{\btzm}[0]{BTZ$_{\rm M}$}
\newcommand{\ads}[1]{{\rm AdS}_{#1}}
\newcommand{\ds}[1]{{\rm dS}_{#1}}
\newcommand{\dS}[1]{{\rm dS}_{#1}}
\newcommand{\eds}[1]{{\rm EdS}_{#1}}
\newcommand{\sph}[1]{{\rm S}^{#1}}
\newcommand{\gn}[0]{G_N}
\newcommand{\SL}[0]{{\rm SL}(2,R)}
\newcommand{\cosm}[0]{R}
\newcommand{\hdim}[0]{\bar{h}}
\newcommand{\bw}[0]{\bar{w}}
\newcommand{\bz}[0]{\bar{z}}
\newcommand{\be}{\begin{equation}}
\newcommand{\ee}{\end{equation}}
\newcommand{\bea}{\begin{eqnarray}}
\newcommand{\eea}{\end{eqnarray}}
\newcommand{\pat}{\partial}
\newcommand{\lp}{\lambda_+}
\newcommand{\bx}{ {\bf x}}
\newcommand{\bk}{{\bf k}}
\newcommand{\bb}{{\bf b}}
\newcommand{\BB}{{\bf B}}
\newcommand{\tp}{\tilde{\phi}}
\hyphenation{Min-kow-ski}

\newcommand{\pa}{\partial}
\newcommand{\eref}[1]{(\ref{#1})}

\def\apr{\alpha'}
\def\str{{str}}
\def\lstr{\ell_\str}
\def\gstr{g_\str}
\def\Mstr{M_\str}
\def\lpl{\ell_{pl}}
\def\Mpl{M_{pl}}
\def\varep{\varepsilon}
\def\del{\nabla}
\def\grad{\nabla}
\def\tr{\hbox{tr}}
\def\perp{\bot}
\def\half{\frac{1}{2}}
\def\p{\partial}
\def\perp{\bot}
\def\eps{\epsilon}

\newcommand{\BC}{\mathbb{C}}
\newcommand{\BR}{\mathbb{R}}
\newcommand{\BZ}{\mathbb{Z}}
\newcommand{\bra}[1]{\langle{#1}|}
\newcommand{\ket}[1]{|{#1}\rangle}
\newcommand{\vev}[1]{\langle{#1}\rangle}
\newcommand{\Real}{\mathfrak{Re}}
\newcommand{\Imag}{\mathfrak{Im}}
\newcommand{\talpha}{{\widetilde{\alpha}}}
\newcommand{\Ham}{{\widehat{H}}}
\newcommand{\al}{\alpha}
\newcommand\x{{\bf x}}
\newcommand\y{{\bf y}}

\def\NPB{{\it Nucl. Phys. }{\bf B}}
\def\PL{{\it Phys. Lett. }}
\def\PRL{{\it Phys. Rev. Lett. }}
\def\PRD{{\it Phys. Rev. }{\bf D}}
\def\CQG{{\it Class. Quantum Grav. }}
\def\JMP{{\it J. Math. Phys. }}
\def\SJNP{{\it Sov. J. Nucl. Phys. }}
\def\SPJ{{\it Sov. Phys. J. }}
\def\JETPL{{\it JETP Lett. }}
\def\TMP{{\it Theor. Math. Phys. }}
\def\IJMPA{{\it Int. J. Mod. Phys. }{\bf A}}
\def\MPL{{\it Mod. Phys. Lett. }}
\def\CMP{{\it Commun. Math. Phys. }}
\def\AP{{\it Ann. Phys. }}
\def\PR{{\it Phys. Rep. }}


\newtheorem{theorem}{Theorem}[section]
\newtheorem{lemma}[theorem]{Lemma}
\newtheorem{proposition}[theorem]{Proposition}
\newtheorem{corollary}[theorem]{Corollary}
\newtheorem{definition}[theorem]{Definition}


\newcommand{\extd}{{\rm d}}

\def\be{ \begin{equation}}
\def\ee{\end{equation}}
\def\bes{\begin{eqnarray}}
\def\ees{\end{eqnarray}}


\title{Modified gravity without  new degrees of freedom}
\author{Laurent Freidel}
\email{lfreidel@perimeterinstitute.ca}
\affiliation{Perimeter Institute for Theoretical Physics, 31 Caroline St. N., Waterloo, N2L 2Y5, Ontario, Canada.}

\begin{abstract}
We show that the new type of ``non-metric'' gravity theories introduced independently by Bengtsson and Krasnov can in fact 
be reexpressed explicitely as a metrical theory coupled to an auxiliary field.
We unravel why such theories possess only one propagating graviton by looking at the quadratic perturbation around a fixed solution. 
And  we give a general construction principle with a new class of example of such modified gravity theories still possessing only two propagating degrees of freedom.

\end{abstract}

\maketitle
\newpage

\section{Introduction }
The Plebanski formulation of general relativity is a very elegant reformulation of the Einstein action and Einstein equations for 4d gravity
 in terms of self dual 
component of the frame field and the corresponding self dual curvature \cite{Plebanski}.
In this formulation one starts from  an $SU(2)$ BF theory
\be
S_{BF}= \int B^{i}\wedge F^{jk}(A)\epsilon_{ijk}
\ee
where $A^{i}_{j}$ is a $SU(2)$ connection and $B^{i}$ an $SU(2)$ valued 2-form field.
In order to obtain the theory of gravity 
 one just  impose an additional ``simplicity constraints'' \cite{Capovilla1,Capovilla2,Reisenberger1}  \be B^{i}\wedge B^{j} \sim \delta^{ij}\label{simp}\ee
 which in turns implies that $F$ is the self dual part of the Riemann tensor associated with a metric reconstructed from the $B$ field.
This formulation has many spin-offs which are potentially essential for our understanding of 
quantum gravity. 
First, the hamiltonian analysis of such an action naturally leads to the formulation of the canonical theory in terms of the Ashtekar variables
\cite{Ashtekar}.
Moreover, this Plebanski formulation of gravity as a constraint $BF$ theory opens the way toward a ``spin foam'' quantisation of 
gravity. See \cite{EPR, FK} for a recent implementation of this program in the context of the SO(4) extension of Plebanski formulation
\cite{DePietri, Reisenberger2}.

In a recent series of works K. Krasnov proposed to study an infinite  class of theories which are natural deformation 
 of the self dual Plebanski formulation of gravity \cite{Krasnov1,Krasnov2}.
 This class of theories is obtained by relaxing the simplicity constraints (\ref{simp}), as we will see more precisely in section \ref{modgrav}.
 They are related to a class of theories studied earlier by I. Bengtsson \cite{Bengtsson,Bengtssonform1,Bengtssonneighbors}. 
 What is remarkable is the fact that this proposal leads to  modification of general relativity which 
 however still contains only two degrees of freedom. This has been demonstrated in an Hamiltonian analysis \cite{BengH, KirillH}.
 When the simplicity constraints are relaxed there seems to be no preferred metrical interpretation of the theory and for this reason this type of theories 
 have been dubbed ``non metric gravity''.
 
 What we  show here is that these theories can in fact be understood as usual  metrical theories. 
 This is achieved in section \ref{effective}  where we reformulate {\it explicitely} the SU(2) BF theory in terms of 
 a gravity theory coupled to an auxiliary  3-dimensional matrix field of unit  determinant. 
 We see in section \ref{modgrav} that the modified theories of Bengtsson-Krasnov just amount in this language to chose an arbitrary potential  for the auxiliary field.
 If one integrate out the matrix field they can be recast as a purely metrical theory which possess a non local action.
 From this perspective it is even more surprising  that such complicated metrical theories have only two degrees of freedom.
 In order to understand what is happening we study the small metric fluctuation and shows in section \ref{pert} that after a non local field redefinition 
 one field behave as a graviton while the other has purely algebraic equation of motion and is therefore not propagating.
 This gives us a general principle allowing us to construct metrical theories which are modification of Einstein gravity but still carry only two degrees of freedom.  
  This principle extends the analysis of \cite{Carroll} where a similar mechanism was studied in the context
  of a scalar (dilatonic) auxiliary field.
  
   Our work overlapp with \cite{Krasnovsame} which appeared 
during the completion of our project and reached some conclusions similar to ours, althought
without giving the details of the effective action.

\section{Urtbanke metric and on-shell connection} \label{metric}

We start from an $SU(2)$ BF theory
\be
S= \int B^{i}\wedge F(A)^{jk}\epsilon_{ijk}
\ee
where $A^{i}_{j}$ is a $SU(2)$ connection and $B^{i}$ an $SU(2)$ valued 2-form field.
This action is quadratic in $A$. We can therefore integrate the connection field. The corresponding equation of motion is given by
\be\label{eq1}
\epsilon^{\alpha \beta \gamma \delta}\nabla^{A}_{\alpha} B^{i}_{\beta \gamma} =0.
\ee
This equation can be solved explicitely \cite{Deser,Halpern,Bengtsson2form} and we will give a description of the solution in the next section.
But before doing so we need to find a new parametrisation of the field $B^{i}$.
The main result we need is the fact known since Urbantke that an $SU(2)$ valued  2-form determines uniquely 
a metric.

More precisely, given an $SU(2)$ valued two form field $B^{i}_{{\mu\nu}}$
 we can define  from it a 4-dimensional  metric $g_{\mu\nu}$ and a 3-dimensional  unimodular symmetric matrix $h^{ij}$.
 Let us start by consider a densitized version of $g,h$ denoted $\tilde{g}, \tilde{h}$ and which are given by 
\be\label{gdef}
 \tilde{g}_{\mu\nu} \equiv\frac1{6} B^{i}_{\mu\alpha} B^{j}_{\beta \gamma}B^{k}_{\delta\nu}\epsilon_{ijk}\epsilon^{\alpha \beta \gamma \delta} ,
\ee
\be \label{hdef}
\tilde{h}^{ij}\equiv  \frac1{8} B^{i}_{\alpha \beta} B^{j}_{\gamma \delta}\epsilon^{\alpha \beta \gamma \delta}.
\ee
$B^{i}$ contains $3\times 6=18$ components, the data 
$h^{ij}, g_{\mu \nu}$ contains $9 + 10 =18 + 1$ components.
This suggests that we can parametrize uniquely the B-field in terms of  these data modulo one relation between $\tilde{g}$ and $\tilde{h}$.
 As we will see this is indeed the case
if $\tilde{g}_{\mu\nu}$ hence   $\tilde{h}^{ij}$ are non degenerate. 
In order to see this, it is convenient to  denote 
\be
\tilde{B}^{i \mu \nu} \equiv \frac12 \epsilon^{\mu \nu \rho \sigma} B^{i}_{\rho \sigma},
\ee
and to use the notation where $B$ and $\tilde{B}$ are treated as $4$ by $4$ matrices, with this notation we 
have 
\be \label{hgdef}
\tilde{g} = \frac16 \epsilon_{ijk}B^{i}\tilde{B}^{j}B^{k}, \quad \tilde{h}^{ij} =\frac14 \mathrm{tr}(B^{i} \tilde{B}^{j}).
\ee
We have the following
 
{\bf{Theorem}} 
{\it
$g_{\mu\nu}$ is a symmetric matrix which is invertible if and only if $\tilde{h}$ is invertible, moreover 
\be\label{const}
\mathrm{det}(\tilde{g})= \mathrm{det}(\tilde{h})^{2}.
\ee
When $\tilde{h}$ is invertible then $B^{i}$ is a two form 
which is self-dual (resp. anti self-dual) with respect to $\tilde{g}$ if $\mathrm{det}(\tilde{h})<0$ 
(resp. if $\mathrm{det}(\tilde{h})>0$) , that is 
\be 
(\tilde{g}\tilde{B}^{i}\tilde{g})_{\mu\nu}= \epsilon   \sqrt{ \tilde{g}}{B}^{i}_{\mu\nu}
\ee
where $\epsilon =-\mathrm{sign}(\mathrm{det}(\tilde{h}))$.
Moreover $\tilde{g}$ is of Euclidean signature if   $\tilde{h}$ is  of Euclidean signature or of Kleinian signature if $\tilde{h}$ is  of Lorentzian signature.
}

Note that since $\tilde{h}$ and $\tilde{g}$ contains one $\epsilon$ tensor in their definition they do not transform as tensors but a densities. It is therefore convenient to introduce the tensor $g_{\mu\nu}$ and scalar $h_{ij}$ as follows
\be
\tilde{g}_{\mu \nu} \equiv \sqrt{g} g_{\mu \nu}, \quad 
\tilde{h}^{ij} \equiv -\epsilon \sqrt{g} h^{ij}.
\ee
The constraint (\ref{const}) imply that 
 $h^{ij}$ is {\it unimodular}, $\mathrm{det}({h})=1$.

The Urbantke metric (\ref{gdef}) can be either Euclidean or Kleinian, so overall this leads to eight different sectors 
which labels: the different signature of $g$, wether $B$ is self or antiself dual with respect to this metric and the overall sign of $B\to -B$.

\subsection{Proof}
This theorem first appeared in \cite{Urbantke}.
We give here a algebraically simple proof of it. 
First lets 
recall that $B$ and $\tilde{B}$ are treated as $4$ by $4$ matrices.
Since $B, \tilde{B}$ are skew symmetric
they satisfy the following key property
\be  \label{Bsym}
B^{i} \tilde{B}^{j}+ B^{j} \tilde{B}^{i}=  \tilde{B}^{j} B^{i}+  \tilde{B}^{i} B^{j}=2 {\tilde{h}^{ij}}  {\mathbf 1}.
\ee
This is easily checked once one notice that the definition of $\tilde{h}^{ij}$ involve a total antisymmetrisation over the $4$ indices of 
a product of $B$s. Thanks to the antisymmetry of $B$ this can be  written as a sum over two terms involving only an antisymmetrisation over 3 indices.
Using this property it is direct to check that 
\be\label{inv}
B^{i}\tilde{B}^{j}B^{k}- B^{k}\tilde{B}^{j}B^{i}= 2\epsilon^{ijk} \tilde{g}.\ee
One first verify that due to (\ref{Bsym}), the LHS is totally antisymmetric in $i,j,k$ the proportionality factor being determined by  the definition (\ref{hgdef}).

We now assume that $\tilde{h}$ is {\it invertible} and denotes $\tilde{h}_{ij}$ its inverse
and defines $\tilde{B}_{i}\equiv \tilde{h}_{ij} \tilde{B}^{j}$.
One first establish from (\ref{Bsym}) that 
\be\label{Bsymcons}
\tilde{B}_{i} {B}^{i} =3 {\mathbf 1}= {B}^{i}\tilde{B}_{i} , \quad 
\tilde{B}_{j} B^{i} \tilde{B}^{j} = - B^{i},\quad
\ee
where repeated index are summed over.
Multiplying (\ref{inv}) on the left  with $\tilde{B}_{j}$ we obtain, after a summation over $j$
and using (\ref{Bsymcons}), that 
 \bea\label{BB}
 \tilde{B}^{j}B^{k}- \tilde{B}^{k}{B}^{j} =  2 \epsilon^{ijk} \tilde{B}_{i} \tilde{g}
\eea
We can  exchange the role of $B$ with $\tilde{B}$ and denote 
\be
\tilde{\tilde{g}}\equiv \frac16 \epsilon_{ijk}\tilde{B}^{i}{B}^{j}\tilde{B}^{k}=
{\mathrm{det}}(\tilde{h})  \frac16\epsilon^{ijk}\tilde{B}_{i}{B}_{j}\tilde{B}_{k}.
\ee
By the same token we obtain identities similar to (\ref{BB})
 \bea
 \tilde{B}^{j}B^{k}- \tilde{B}^{k}{B}^{j} =  2\epsilon^{ijk} \tilde{\tilde{g}} {B}_{i},
\eea
hence 
\bea
\tilde{B}^{i}B^{j}=\tilde{h}^{ij}+ \epsilon^{ijk} \tilde{B}_{k} \tilde{g}, 
  \quad\mathrm{and}\quad \tilde{B}_{i} \tilde{g}=\tilde{\tilde{g}} {B}_{i}.
\eea
We will also use the transpose of the first identity which reads 
${B}^{i}\tilde{B}^{j}=\tilde{h}^{ij}+ \epsilon^{ijk} \tilde{g} \tilde{B}_{k}.$
From this we  we can eventually check by a direct computation that 
 \bea
 \tilde{g}\tilde{\tilde{g}}&=&
\frac14 {\mathrm{det}}(\tilde{h})(B^{1}\tilde{B}^{2}B^{3}- B^{3}\tilde{B}^{2}B^{1})
 (\tilde{B}_{1}{B}_{2}\tilde{B}_{3}-\tilde{B}_{3}{B}_{2}\tilde{B}_{1})\\
 &=&- {\mathrm{det}}(\tilde{h})\mathbf{1}
 \eea
 It is easier to first assume that $\tilde{h}$ is diagonal 
 to directly show this property (since in this case 
 $\tilde{B}_{1}{B}_{2}\tilde{B}_{3} = -\tilde{B}_{3}{B}_{2}\tilde{B}_{1}$, $B^{3}\tilde{B}_{3}= \mathbf{1}$) and then that the equality 
 is independent under conjugation of $\tilde{h}$.
 
 Therefore, one sees that $\tilde{g}$ is invertible if and only if $\tilde{h}$ is invertible
 and that we have the self duality relation 
 \be\label{sd2}
 \tilde{g}\tilde{B}^{i}\tilde{g}=-  {\mathrm{det}}(\tilde{h}) B^{i}.
 \ee
 Applying this self duality relation twice we obtain the identity
 $${\mathrm{det}}(\tilde{g})={\mathrm{det}}(\tilde{h})^{2}.$$
 There are different discrete sectors which are related to each other either by a change of sign of $B\to -B$ in which case $\tilde{g}\to -\tilde{g}$ and $\tilde{h}\to \tilde{h}$ or by a change of orientation $\epsilon_{\mu\nu\rho\sigma} \to - \epsilon_{\mu\nu\rho\sigma}$  in which case  $\tilde{g}\to -\tilde{g}$ and $\tilde{h}\to -\tilde{h}$.

It will be convenient to introduce the frame field $e^{A}_{\mu}$ associated with the metric $g$
\be
g_{\mu\nu} =\eta_{AB} e^{A}_{\mu} e^{B}_{\nu}.
\ee
where $\eta_{AB}$ is a diagonal metric, whose eigenvalue are $\pm 1$ and  which is either Euclidian  or Kleinian (of signature $(--++)$), since 
$\mathrm{det}(g)>0$. The frame field orientation is chosen such that 
$\mathrm{det}(e) =\sqrt{{g}}$.
Modulo  the discrete degeneracy $B\to -B$ described above we can always choose  
to be in the sector where $\eta_{00} =+1$  and we define the sign of $\mathrm{det}(\tilde{h})$ to be $-\epsilon$.
The other sectors are related either by 
by a change of global sign of $\eta_{AB}$.
The identity (\ref{sd2})  tells us that we can expand the $B$ field in terms of self or anti-self dual components
 \be 
 B^{i} = b^{i}_{a} \Sigma^{ a}_{\epsilon}(e), \quad \mathrm{with} \quad \Sigma^{a}_{\epsilon\mu \nu}(e) = \sigma^{a}_{\epsilon AB} e_{\mu}^{A}e_{\nu}^{B}
 \ee
where 
\be
\sigma^{a}_{\epsilon AB}= (\delta^{0}_{A}\delta^{a}_{B}-\delta^{0}_{B}\delta^{a}_{A})+ \epsilon \epsilon^{0aA'B'}\eta_{A'A}\eta_{B'B}
\ee
$\sigma_{+}^{a}$ is the 't hooft symbol which projects onto the space of self dual tensors.
A direct computation shows that 
\be
\frac14 \sigma^{a}_{\epsilon AB} \tilde{\sigma}^{b BA}_\epsilon = -\epsilon  \eta^{ab},\quad 
\frac16 \epsilon_{abc} (\sigma^{a}_{\epsilon} \tilde{\sigma}^{b}_{\epsilon} \sigma^{c}_{\epsilon})_{AB} = \epsilon  \eta_{AB}.
\ee
where we have use that in the chosen sector  $\eta_{00}=1$ and $\eta_{11}\eta_{22}\eta_{33}=1$.

This implies that the signature of $h$ is Lorentzian iff the signature of $g$ is Kleinian as claimed.
Moreover from these relation one sees that obtains that the field $b$ is a three dimension frame  frame field for the metric $h^{ij}$
and  is unimodular:
\be
\tilde{h}^{ij} = - \epsilon \sqrt{g} \,  b_{a}^{i} \eta^{ab} b_{b}^{j},\quad \quad
\mathrm{det}{(b)}= 1.
\ee
$\Box$

\subsection{Connection}
We can now summarise the results of the previous section as follows:
 A general  SU(2) valued two form $B^{i}$ can be equivalentely described in terms of 
 a four dimensional  frame field $e_{\mu}^{A}$ determining a spacetime metric 
 $g_{\mu\nu} = \eta_{AB}e^{A}_{\mu}e^{B}_{\nu}$,
 and a 3 dimensional unimodular ($\mathrm{det}(b)=1$)  ``internal'' frame field   $b^{i}_{a}$ determining a unimodular scalar metric
 $h^{ij} = b_{a}^{i}\eta^{ab}b_{b}^{j}$.
 There are two main sector in which either  $\eta_{AB}= (\eta_{00},\eta_{ab}) = \mathrm{diag}(++++)$ or 
  $(\eta_{00},\eta_{ab}) =\mathrm{diag}(++--)$.
  In  each case the $B^{i}$ field can be uniquely reconstructed from these data modulo a  fourfold discrete ambiguity parametrised by  two signs $\tilde{\epsilon},\epsilon = \pm 1$:
 \be
 B^{i}= \tilde{\epsilon}  b_{a}^{i} \Sigma^{a}_{\epsilon}(e).
 \ee
 with $\Sigma^{a}_{\pm}(e)\equiv 2\left(e^{0}\wedge e^{a} \pm \epsilon^{abc} e_{b}\wedge e_{c}\right)$, a basis of self (or anti self) dual bivectors which satisfies
 \be\label{SS}
 \Sigma^{a}_{\epsilon}\tilde{\Sigma}^{b}_{\epsilon}
 = -  \sqrt{g}\left(  \epsilon \eta^{ab} + \epsilon^{abc} \eta_{c\bar{c}} \Sigma^{\bar{c}}_{\epsilon} g^{-1}\right),\quad \quad 
 g\tilde{\Sigma}^{a}_{\epsilon} g  = \epsilon \sqrt{g} {\Sigma}^{a}_{\epsilon}.
 \ee
  In each sector the data  $\epsilon,\tilde{\epsilon}, e^{A}, b_{a}$ is uniquely determined by 
  $B^{i}$ modulo  an SO(4) (resp. SO(2,2)) rotation acting on $e^{A}$ and an induced SO(3)  (resp. SO(2,1)) self dual rotation acting on $b_{a}$.

The purpose of this section is to give a solution of  equation (\ref{eq1}). 
Such a solution has been provided a long time ago \cite{Deser,Halpern} but we will need 
to   give here an independent derivation of this solution.
The advantages of our derivation is that we can express the connection as a linear sum of a gravitational spin connection 
and an additional one form. This decomposition will be essential for us in order to 
construct the effective gravitational description of SU(2) $BF$ theory.

In order to solve (\ref{eq1}) one first denote by  $b^{a}_{i}$
 the inverse of the three dimensional internal frame field $b_{a}^{i}$ (indices $i,j,k$ denotes SU(2) indices carried by the 
 $B$ field while indices $a,b,c$ are ``internal ''  SU(2) indices.
And we  introduce the following connection
\be
\omega^{a}{}_{b}= b^{a}_{i}A^{i}_{j}b^{j}_{b} + b^{a}_{i} d b^{i}_{b} =
(b^{{-1}}Ab+b^{{-1}}db)^{a}_{b}
\ee
This connection is such that
\be 
\label{eq2}d_{A} B^{i}=\tilde{\epsilon}\,  b^{i}_{a} d_{\omega}\Sigma^{a}_{\epsilon}(e) =0.
\ee
Thus $\omega$ satisfy a  condition of zero torsion.
If $\omega$ was satisfying the additional  metricity condition 
\be
0=d_{\omega}\eta_{ab} 
\ee then we could easily solve  (\ref{eq2})  since 
 $\omega$ would just be the self dual part of the spin connection 
associated with $e$.
More precisely\footnote{Explicitely this gives 
\be
 \gamma_{\nu }^{a}{}_{b} = \eta_{b\bar{b}}\epsilon^{a\bar{b}c}\left( \partial_{\mu} (\tilde{\Sigma}_{c}g)^{\mu}_{\nu}+ \epsilon \epsilon_{abc} \partial_{\mu} \tilde{\Sigma}^{a \mu\rho}\Sigma^{b}_{\rho\nu}\right)\ee
}, given the spin connection $\gamma$ solution of 
$de^{A} + \gamma^{A}{}_{B}\wedge e^{B}=0$ we can define its self-dual (or anti self-dual) projection.
\be 
{\gamma^{a}_{\epsilon}}_{b}(e) \equiv \sigma^{c}_{\epsilon AB} \gamma^{A}{}_{B'} \eta^{B' B} \eta^{aa'}\epsilon_{a'bc}
= \sigma^{c}_{\epsilon AB} \gamma^{AB}\epsilon^{a}{}_{bc}.
\ee

In our case the connection $\omega$ is non metric indeed 
what $\omega$ preserve is not  $\eta_{ab}$ but 
\be
b_{ab} \equiv b_{a}^{i}\eta_{ij}b_{b}^{j}.
\ee
where $\eta_{ij} =\mathrm{diag}(+++)$ is the SU(2) metric.
Indeed 
\be \label{eq3}
0=d_{A} \eta_{ij}= d_{A}(b_{i}^{a}b_{ab} b_{j}^{b})= b_{i}^{a} b_{j}^{b}(d_{\omega}b_{ab})=0.
\ee

One has to be careful in our  manipulation of indices, since we have two natural metric on the space of internal indices $a$, the flat metric 
$\eta_{ab}$ preserved by the 
spin connection $\gamma_{\epsilon}$ and the frame metric $b_{ab}$ preserved by $\omega$.
In the following we  denote by
$\hat{b}^{ab}$ the inverse of $b_{ab}$ and $\eta^{ab}$ the inverse of $\eta_{ab}$.
Unless explicitely specify we {\it do not}
 use a convention where upper indices are raised
 with respect to $b^{ab}$ or $\eta^{ab}$. The only exception is the epsilon tensor since both metric are unimodular, we have $$\epsilon^{abc}\equiv \eta^{a \bar{a}} \eta^{b \bar{b}} \eta^{c \bar{c}} \epsilon_{abc}=b^{a \bar{a}} b^{b \bar{b}} b^{c \bar{c}} \epsilon_{abc}.$$ 
Moreover we denote 
\be
d_{\omega} \equiv dx^{\mu} D_{\mu}, \quad d_{\gamma} \equiv dx^{\mu} \nabla_{\mu}.
\ee
We want to show that there is a unique solution of the torsion + non metricity equations (\ref{eq2}, \ref{eq3}):
\be
D_{\mu} \tilde{\Sigma}^{a\mu\nu} =0, \quad  \partial_{\mu}b_{ab} =2 \omega_{\mu}{}^{c}{}_{(a} b_{b) c}.
\ee
where $(ab)$ means symmetrisation.

Suppose that $\omega = \gamma + \rho$ where $\gamma$ is the spin connection and $\rho$ is a one form.
The zero torsion condition then reads
\be\label{eq4}
\rho_{\mu}{}^{a}{}_{b}\tilde{\Sigma}^{b\mu\nu} =0
\ee
since $0=d_{\gamma}\Sigma^{a}(e)$ by definition.
The symmetric part of $\rho$ is determined by the non-metricity equation, hence
\be
 \rho_{\mu ab} \equiv b_{ac} \rho_{\mu}{}^{c}{}_{b}, \quad \rho_{\mu(ab)} =\frac12 \nabla_{\mu}b_{ab},\quad \rho_{\mu[ab]} \equiv \epsilon_{abc }\rho_{\mu}^{c} .
\ee
The zero torsion equation can then be written as 
\be 
\rho_{\mu}^{c} \epsilon_{abc }\tilde{\Sigma}^{b\mu\nu} =
\rho_{\mu}^{c} \tilde{\Sigma}^{\mu\nu}_{ca}= 
-\frac12 \nabla_{\mu}b_{ab}\tilde{\Sigma}^{b\mu\nu}
\ee
where we have denoted $\tilde{\Sigma}^{\mu\nu}_{ca}\equiv  \epsilon_{bca }\tilde{\Sigma}^{b\mu\nu} $.
One multiply both sides of this equation with $(\Sigma ^{\bar{c}}\tilde{\Sigma}^{a} g)_{\nu \bar{\mu}}$ and use the identity 
following from (\ref{SS}) and its transpose, one gets
\be
\tilde{\Sigma}^{a}_{\epsilon}\Sigma ^{b}_{\epsilon}\tilde{\Sigma}^{c}_{\epsilon} g 
=\mathrm{det}(g)\left\{\epsilon \epsilon^{abc} + g^{-1} \left(  \eta^{ac} \Sigma^{b}_{\epsilon} -2 \eta^{b(a}\Sigma^{c)}_{\epsilon} \right)\right\}.
\ee
Therefore 
\bea
\rho_{{\mu}}^{{c}} &=&  \frac{\epsilon}{4\sqrt{g}^{2}} \nabla_{\bar{\mu}}b_{ab}(\tilde{\Sigma}^{b}_{\epsilon}\Sigma ^{{c}}_{\epsilon}\tilde{\Sigma}^{{a}}_{\epsilon} g)^{\bar{\mu}}{}_{{\mu}}\\
&=& -\frac{\epsilon}{2} (\nabla_{\bar{\mu}}b_{ab}) 
(g^{-1}[ {\Sigma}^{(a}_{\epsilon}\eta^{b)c} - \frac{1}2\eta^{ab}{\Sigma}^{{c}}_{\epsilon}] )^{\bar{\mu}}{}_{{\mu}}\\
&=& \frac{\epsilon}{2} 
[ {\Sigma}^{(a}_{\epsilon}\eta^{b)c} - \frac{1}2\eta^{ab}{\Sigma}^{{c}}_{\epsilon}] _{\mu \nu} (\nabla^{\nu}b_{ab})
\eea
Thus $\omega = \gamma_{\epsilon}(e) + \rho$ with
\be
\rho_{\mu ab} =\frac1{2} (\nabla_{{\mu}}b_{{a}{b}}) +\frac{\epsilon}{2}  \epsilon_{cab}
[ {\Sigma}^{(\bar{a}}_{\epsilon}\eta^{\bar{b})c} - \frac12\eta^{\bar{a}\bar{b}}{\Sigma}^{{c}}_{\epsilon}] _{\mu \nu} (\nabla^{\nu}b_{\bar{a}\bar{b}}).
\ee
 is the unique solution of (\ref{eq2}).

\section{The effective BF action}\label{effective}

We can now put back into the BF action the solution we just found. 
In order to do so and in order to avoid a notational cluttering  we chose to work for the rest of the paper 
in the self dual sector where $\epsilon=+1$
and $\tilde{\epsilon}=+1$. And we denote from now on $\Sigma^{a} \equiv \Sigma_{+}^{a}$.
The other sector can be worked out in exactly the same way

After integration over the connection  $A$ the $SU(2)$ BF action simply becomes 
\bea\nonumber
S_{BF} &=&\int b^{i}_{a}\Sigma^{a}(e)  F^{jk}(b\omega b^{-1}+dbb^{{-1}}) \epsilon_{ijk}
=   \int \Sigma^{a}(e)\wedge  F^{b}{}_{c}(\omega) b^{i}_{a}b^{j}_{b}b^{c}_{k}\epsilon_{ij}{}^{k}\\
&=&   \int \Sigma^{a}(e) \wedge F^{b}{}_{c}(\omega)\epsilon_{ab\bar{c}}\hat{b}^{\bar{c}c}
\eea
Now since $\omega =\gamma(e) +\rho$ 
we can expand $S_{BF}= S_{1}+ S_{2} +S_{3}$ as a sum of three terms
\bea
S_{1} &=& \int    \Sigma^{a}(e) \wedge F^{b}{}_{c}(\gamma(e))\epsilon_{ab\bar{c}}\hat{b}^{\bar{c}c}\\
&=&  \int\left[ \Sigma^{{a}}(e) \wedge F_{a}(\gamma(e))\, \eta_{bc}\hat{b}^{bc} -  \Sigma^{{a}}(e) \wedge F_{b}(\gamma(e)) \,\eta_{ac}\hat{b}^{cb}\right]
\\
&=&2  \int \sqrt{g} R_{ab}(e) [\hat{b}^{ab}- \eta^{ab} \,\eta_{\bar{a}\bar{b}}\hat{b}^{\bar{a}\bar{b}} ]
\eea
where $F_{a}\equiv \frac12\epsilon_{abc} \eta^{c\bar{c}} F^{b}{}_{\bar{c}}$ and 
 $R_{ab}(e)$ is the self-dual   part of the Riemman tensor:
 $2 F_{b\mu\nu}\equiv  R_{bc} \Sigma^{c}_{\mu\nu}$.

The second term is given by 
\bea
S_{2}&=& \int   \epsilon_{ab\bar{c}}\, \Sigma^{a}(e) \wedge d_{\gamma} \rho^{b}{}_{c}\hat{b}^{c\bar{c}}
=\int   \epsilon_{ab\bar{c}} \Sigma^{a}(e) \wedge \rho^{b}{}_{c} \wedge d_{\gamma}\hat{b}^{c\bar{c}}\\
&=& -2 \int   \epsilon_{ab\bar{c}} \Sigma^{a}(e) \wedge \rho^{b}{}_{c} \wedge \rho^{(c}{}_{d}    \hat{b}^{\bar{c}) d}
\eea
where we have integrated by part in the first equality and use the metricity condition $d_{\gamma+\rho}b_{ab}=0$ in the second.
Finally 
\be
S_{3}= \int \epsilon_{ab\bar{c}}\, \Sigma^{a}(e) \wedge \rho^{b}{}_{c} \wedge\rho^{c}{}_{d}\,\hat{b}^{\bar{c}d}.
\ee
In the rest of the section it is convenient to use a notation where internal indices of $\rho$ are raised with $\hat{b}$, i-e 
$ \rho^{ab} \equiv \rho^{a}{}_{c} \hat{b}^{cb}$,
$\rho_{ab} \equiv {b}_{ac}\rho^{c}{}_{b} $.
Thus
\bea
\tilde{S}=S_{2}+S_{3}& =&-  \int \epsilon_{ab\bar{c}}\, \Sigma^{a}(e) \wedge \rho^{b}{}_{c} \wedge\rho^{\bar{c}c}\\
& =& -  \int \epsilon_{ab{c}}\, \Sigma^{a}(e) \wedge \rho^{b\bar{b}} \wedge\rho^{c\bar{c}} \,b_{\bar{b}\bar{c}}
\eea
We can further simplify this expression by expanding $\rho$ in terms of its symmetric and skew symmetric parts.

Let us first look at 
\bea
& &\int  \epsilon_{abc} \Sigma^{a}\wedge \rho^{[b\bar{b}]}\wedge \rho^{c\bar{c}}\, b_{\bar{b}\bar{c}}
= \int  \epsilon_{abc}\epsilon^{b'b\bar{b}} \Sigma^{a}\wedge \rho_{b'}\wedge \rho^{c\bar{c}} \,b_{\bar{b}\bar{c}}\\
&=&- \int \Sigma^{a}\wedge \rho^{b}\wedge \rho_{ba} +\int \Sigma^{c}\wedge \rho_{c}\wedge \rho^{ba} b_{ab} =0
\eea
The first term is equal to zero since our defining equation for $\rho$ is 
$\rho_{ab}\wedge \Sigma^{b}=0$ and the second term is equal to zero also because
\be
 \rho^{ba} b_{ab} = \frac12 d_{\gamma} b^{ab} b_{ab}= \sqrt{\mathrm{det}(b)}^{-1}d_{\gamma} \sqrt{\mathrm{det}(b)}=0
 \ee
 since ${\mathrm{det}(b)}=1$.
 Thus 
 \bea
\tilde{S}=\tilde{S}_{1}+\tilde{S}_{2}= -\int  \epsilon_{abc} \Sigma^{a}\wedge \rho^{[b\bar{b}]}\wedge \rho^{(c\bar{c})}\, b_{\bar{b}\bar{c}} -
\int  \epsilon_{abc} \Sigma^{a}\wedge \rho^{(b\bar{b})}\wedge \rho^{(c\bar{c})}\, b_{\bar{b}\bar{c}}\\
= \frac12 \int \rho^{b}\wedge \Sigma^{a}\wedge d_{\gamma} b_{ab}
-\frac14 \int  \epsilon_{abc} \,d_\gamma \hat{b}^{b\bar{b}}\wedge \Sigma^{a}\wedge d_\gamma \hat{b}^{c\bar{c}}\, b_{\bar{b}\bar{c}} 
\eea
Lets focus on the first term, using the notation of the previous section we can write it as 
\bea
\tilde{S}_{1} &=& \int  \rho_{\mu}^{b} \tilde{\Sigma}^{a\mu\nu}\nabla_{\nu}b_{ba}\equiv
\int  \rho^{ b} \tilde{\Sigma}^{a}(\nabla b_{ab}) =-\int (\nabla b_{ab}) \tilde{\Sigma}^{a} \rho^{ b} \\
&=&-\frac{1}{2} \int (\nabla b_{ab})\left[ \tilde{\Sigma}^{a} \Sigma^{\bar{a}} \eta^{b\bar{b}}- \frac12 \eta^{\bar{a} \bar{b}}\tilde{\Sigma}^{a} \Sigma^{b} \right] g^{-1}(\nabla b_{\bar{a}\bar{b}})\nonumber \\
&=&\frac12 \int (\nabla b_{ab})\left[\sqrt{g} (\eta^{\bar{a}a}\eta^{b\bar{b}} - \frac12\eta^{ab} \eta^{\bar{a}\bar{b}})g^{-1}
+ (\eta\tilde{\Sigma})^{a\bar{a}}\eta^{b\bar{b}} \right] (\nabla b_{\bar{a}\bar{b}})\nonumber \\
\eea
where $(\eta \tilde{\Sigma})^{ab}\equiv 
\epsilon^{ab\bar{c}} \eta_{\bar{c}c} \tilde{\Sigma}^{c}$

We can similarly evaluate the second term
\bea
\tilde{S}_{2} &=& - \frac12 \int (\nabla \hat{b}^{b\bar{b}}) \epsilon_{abc}\tilde{\Sigma}^{a} (\nabla \hat{b}^{c\bar{c}})b_{\bar{b}\bar{c}} \\
&=& - \frac12 \int (\nabla b_{ab}) (b\tilde{\Sigma})^{a\bar{a}} \hat{b}^{b\bar{b}}(\nabla b_{\bar{a} \bar{b}}). \\
\eea
with  $(b\Sigma)_{ab}\equiv 
\epsilon^{ab\bar{c}} b_{\bar{c}c} {\Sigma}^{c}$.
Therefore overall one gets
\be\label{baction}
\tilde{S}=  \frac12 \int (\nabla b_{ab})\left[{\sqrt{g}} (\eta^{a\bar{a}}\eta^{b\bar{b}} - \frac12\eta^{ab} \eta^{\bar{a}\bar{b}})g^{-1}
+ ((\eta\tilde{\Sigma})^{a\bar{a}}\eta^{b\bar{b}}- (b\tilde{\Sigma})^{a\bar{a}} \hat{b}^{b\bar{b}}) \right] (\nabla b_{\bar{a}\bar{b}}).
\ee
the first term is a kinetic term purely quadratic in  $b_{ab}$, the second term is an interaction term 
between $b$ and the metric
which vanishes when $b^{ab}=\eta^{ab}$.
The form of the kinetic term for $b_{ab}$ shows that this field behave like a minimally coupled field. Since $\mathrm{det}(b)=1$ we can
 express the entire action purely in terms of $b_{ab}$.
 One also use a notation where indices on b are raised with respect to the flat metric $\eta$, i-e $b^{a}_{b}\equiv \eta^{a\bar{a}}b_{\bar{a}b}$, $b^{ab}
 \equiv \eta^{a\bar{a}}b_{\bar{a}\bar{b}}\eta^{\bar{b}b}$. Beware that the inverse metric  $\hat{b}^{ab} \neq {b}^{ab}$, it can however be expressed as a quadratic function of $b$
 \be
 \hat{b}^{ab} = b^{ac}b_{c}^{b} - b b^{ab} -\frac12\left(b_{d}^{c}b_{c}^{d} - b^{2}\right) \eta^{ab},\quad \quad b\equiv b^{a}_{a}.
 \ee 
 The SU(2) BF action can be written as a function of the metric $g$ and scalar field
 \bea \label{SBF2}
 S_{BF}(g,b) &=&2\int \sqrt{g} R_{ab}^{\epsilon}\left( b^{ac}b_{c}^{b} - b b^{ab}\right) +\sqrt{g}g^{\mu\nu}\left(\nabla_{\mu}b_{a}^{b}\nabla_{\nu}b_{b}^{a} -\frac12 \nabla_{\mu}b\nabla_{\nu}b
 \right) \nonumber\\
 & & +  \int \sqrt{g}\,\nabla^{\mu}b_{ab} \left((\eta {\Sigma})_{\mu\nu}^{a\bar{a}}\eta^{b\bar{b}}-(b{\Sigma})_{\mu\nu}^{a\bar{a}}\hat{b}^{b\bar{b}}\right)\nabla^{\nu}b_{\bar{a}\bar{b}}.  \eea
 
 It is now easy to see that when $b_{ab}=\eta_{ab}$ the theory reduces to Einstein gravity.
 Indeed the relationship between the self dual curvature tensor  given here and the usual Riemman tensor is 
 \be
R_{\mu \nu}{}^{\alpha \beta} (\eta\tilde{\Sigma}_{a})^{\mu \nu} (\eta \Sigma_{b})_{ \alpha \beta} = 4\sqrt{g} R_{ab},\quad R = R_{\mu \nu}{}^{\mu \nu} = 2 R_{ab} \eta^{ab}
\ee
thus 
\be
S(g_{\mu\nu},\eta_{ab}) = -2 \int \sqrt{g} R(g).
\ee

\section{Modified gravity}\label{modgrav}

  In \cite{Krasnov1,Krasnov2} a general class of gravity theories were obtained by adding to the $BF$ action a term of the form
  \be\label{interact}
 S_{int}= \int (\Psi_{ij} -\eta_{ij} \Lambda(\Psi)) B^{i}\wedge B^{j}
  \ee
where $\Psi_{ij}$ is a traceles symmetric $3\times 3$ matrix and $\Lambda$ is a 
function of $\Psi$ invariant under congugation that is it is a function of 2 variables $x\equiv\frac12\tr(\Psi^{2}), 
y\equiv \frac13 \tr(\Psi^{3})= \mathrm{det}(\Psi)$
since $\Psi$ satisfy the caracteristic equation
$\Psi(\Psi^{2}-\frac12{\tr(\Psi^{2})}) =\det(\Psi).$ 

This can be written in terms of the metric and $h$ field as 
\be
S_{int}= -4 \int \sqrt{g} (\Psi_{ij} -\eta_{ij} \Lambda(\Psi)) h^{ij}.
  \ee
Since the dependence on $\Psi$ is purely algebraic and does not contain derivative 
 we can be integrated out. The equations of motion read
 \be
 \frac{\partial{\Lambda(\Psi)}}{\partial{\Psi_{ij}}} = H^{ij}, \quad {\mathrm{with}}\quad   H^{ij} \equiv \frac{h^{ij}}{\tr(h)}- \frac{\eta^{ij}}{3}
 \ee
 $H$ is symmetric traceless.
 Therefore the action evaluated on-shell leads to a potential for $h^{ij}$
\be
S_{int}= -4  \int \sqrt{g}\, V(h^{ij}) = 4\int V(\tilde{h}^{ij}).
\ee
where $V(h)$ is an homogeneous function of $h$ of degree one $h^{ij}\frac{\partial{V}}{\partial{h^{ij}}}= V$ which is invariant under conjugation $V(khk^{-1})=V(h)$. This is essentially the Legendre transform  of $\Lambda$, that is 
\be 
V(h) = \tr(h) (H^{ij}\Psi_{ij} -F(\Psi)).
\ee
Conversely, if one start from an arbitrary potential $V$ we can define 
$$\tilde{\Lambda}(H) \equiv \frac{V(h)}{\tr(h)}= V\left(\frac{\eta}{3} + H\right)$$ and reconstruct the function $\Lambda$ entering (\ref{interact}) as   the Legendre transform
\be
\Lambda(\Psi) = H^{ij}\Psi_{ij} -\tilde{\Lambda}(H).
\ee

Note that here we have expressed the potential in terms of $\tilde{h}^{ij}=\frac14 \tr(B^{i}\tilde{B}^{j})=  \sqrt{g} b^{i}_{a}\eta^{ab}b^{j}_{b}$ wherehas we have expressed our action in terms of $ b_{ab} =b^{i}_{a}\eta_{ij}b^{j}_{b}$. 
The two formulations are related since $V$ is a homogeneous function of 
$\tr(\tilde{h})=-\sqrt{g} \tr({h})$, $(\tr(\tilde{h})^{-1})^{-1}= \sqrt{g} \tr({h}^{-1})^{-1}$, and  $|\mathrm{det}(\tilde{h})|^{\frac13}=\sqrt{g}$. Since $\mathrm{det}({b})=1$
we can express the potential as an arbitrary function of 
$b_{ab}\eta^{ab}=\tr(h)$ and $\hat{b}^{ab}{\eta}_{ab}=\tr(h^{-1})$.

Moreover, as we have seen in the previous section the $b_{ab}$ behave as a field of mass dimension $1$
so for dimensional reason the potential term should involve a mass scale.
For instance 
if one look to a quadratic potential 
\be
\Lambda(\Psi) = \Lambda +\frac{1}{2 M^{2}} \tr(\Psi^{2}), \quad \tilde{\Lambda}(H) = -\Lambda + \frac{M^{2}}{2} \tr(H^{2}).
\ee
In the limit $M^{2} \to \infty $ the potential term forces $H=0$ hence $h^{ij}=\eta^{ij}$ or $b_{ab}=\eta_{ab}$
and we recover the 
 case of usual gravity with a cosmological constant.

If one choose an arbitrary potential $V(b)$ one therefore obtain an infinite  family of deformation 
of classical general relativity.
In the original references these deformation where dubbed ``non-metrical''. Our analysis shows however that 
such deformation can be given a  purely metrical interpretation: Suppose that we integrate the $b$ field out by solving its 
equation of motion the theory that we obtain is an effective  theory which depends only on a metric and which is invariant under diffeomorphism.
For a generic potential this effective theory is {\it not} general relativity, it is a theory that contains an arbitrary number of higher derivative and curvature terms as we will see more precisely in the next section.
This is in general a non local theory of the metric.

From this perspective, what is remarkable is the statement that the theory still possess only two degree of freedom 
even when the potential $V$ is arbitrary. This has been shown in the Hamiltonian context using a 
canonical analysis \cite{BengH, KirillH} which generalises the original Ashtekar analysis \cite{Ashtekar}.
We would like to give now an understanding of this essential property purely from a metrical point of view. 

\section{Perturbation theory}\label{pert}
In order to understand what are the local degrees of freedom  associated with 
 the theory associated with a a non trivial potential $V$ we look at the fluctuations around a background solution.
 We will assume that the potential $V$ is chosen such that $b_{ab}=\eta_{ab}$ is still a vacuum solution.
 In order to  linearised the action.
 we introduce the parameters
 $e_{\mu}^{A} = \delta_{\mu}^{A} + \bar{e}_{\mu}^{A}$ thus $g_{\mu \nu} =\eta_{\mu \nu} + 2 e_{(\mu \nu)}\equiv \eta_{\mu \nu} +h_{\mu \nu}$ and 
  $b_{ab}=\eta_{ab} + \bar{b}_{ab} $ , $\hat{b}^{ab} =\eta^{ab} - \bar{b}^{ab} +\cdots $ where $\bar{b}$ is traceless
  and all indices are raised and lowered with $\eta_{ab}$.
 The only term which is non trivial to linearise is the first term in (\ref{SBF2}). Expanding this term in powers of $\bar{b}$
 one obtains
 \be
 S_{1}= 2 \int \sqrt{g} R_{ab} (b^{ab} - \eta^{ab} b)
 \sim  - 2\int  \sqrt{g} R(g) + \frac12 \int   R_{\mu \nu}{}^{\rho \sigma}(g) (\eta\tilde{\Sigma}_{a})^{\mu \nu} (\eta \Sigma_{b})_{\rho \sigma}\, \bar{b}^{ab}
 \ee
To developp the perturbation theory we start from the expression of the linearised spin connection
\be
\omega_{\mu}^{AB}\delta_{A\nu}\delta_{B\rho} \equiv
\omega_{\mu \nu \rho} =
 \partial_{\mu} e_{[\nu \rho]} -\partial_{\nu } e_{(\mu \rho)} + \partial_{\rho} e_{(\mu \nu)}
\ee
The Einstein action at quadratic order is given by the Pauli-Fierz action $- 2\int  \sqrt{g} R(g)\sim S_{PF}(h)$ where
\bea
 S_{PF}(h)  &=& - 2\int \left(\omega_{\mu}{}^{\mu \rho}\omega^{\nu}_{\mu \rho} - \omega_{\mu \nu \rho}\omega^{\nu \mu \rho}\right)\\
& = &\frac12 \int \left( \partial_{\rho}h_{\mu \nu}\partial^{\rho}h^{\mu \nu}
-2 (\partial h)_{\rho} (\partial h)^{\rho} +2 (\partial h)_{\rho} \partial^{\rho}h - \partial_{\rho} h \partial^{\rho}h  \right),
\eea
 with $(\partial h)_{\rho} \equiv \partial^{\mu} h_{\mu \rho}$ and 
 $h\equiv h_{\mu \nu} \eta^{\mu \nu}$.

The coupling term is given by 
\bea 
\frac12 \int   R_{\mu \nu}{}^{\rho \sigma}(g) \tilde{\Sigma}_{a}^{\mu \nu}\Sigma_{b\rho \sigma} \, \bar{b}^{ab} =
- \int \partial_{\mu} \partial_{\rho } h_{\nu \sigma}\sigma_{a}^{\mu \nu}\sigma_{b}^{\rho\sigma}\, \bar{b}^{ab} =
 - \int h_{\mu \nu } \partial^{\mu}_{a}\partial^{\nu}_{b} \bar{b}^{ab} 
\eea
where $\sigma_{a}^{\mu \nu} $ is (twice) the t'hooft tensor: $\sigma_{a}^{\mu \nu} B_{\mu \nu} =2 (B_{0a} + \tilde{B}^{0a})$ and 
$\partial^{\mu}_{a} \equiv \sigma_{a}^{\mu \nu}\partial_{\nu}$.
These derivatives satisfy the identity
\be
\partial_{\mu}\partial^{\mu}_{a} = 0,\,\quad 
\partial^{\mu}_{a}\eta_{\mu \nu}\partial^{\nu}_{b} = -\Box \delta_{ab},\quad
\ee
and  the tensor 
\be
b^{\mu \nu} \equiv \frac1{\Box} \partial^{\mu}_{a}\partial^{\nu}_{b} \bar{b}^{ab} 
\ee
is tranverse and traceless
\be
\partial_{\mu} b^{\mu \nu} =0,\quad \eta_{\mu \nu} b^{\mu \nu} = - \delta_{ab} \bar{b}^{ab} =0.
\ee
Therefore if one define the tensor 
\be
\hat{h}_{\mu \nu} = h_{\mu \nu } + b_{\mu \nu}
\ee
we can give a canonical form to the quadratic action 
\be
S_{1} = S_{PF}(\hat{h}) + \frac12 \bar{b}^{ab} \Box \bar{b}_{ab}.
\ee
 The key point is that the quadratic kinetic  term in $\bar{b}$ is exactly cancelled by 
 the contribution from the quadratic expansion of (\ref{baction}). 
 This means that the field  $\bar{b}$ is non dynamical since the total `modified gravity' action is given 
 at quadratic level by
 \bea
 S &=& S_{PF}({h}) - h_{\mu \nu}\Box b^{\mu \nu} +\frac12 \partial_{\mu}\bar{b}^{ab}\partial^\mu\bar{b}_{ab}- \frac{M^{2}}{2} \bar{b}^{ab}\bar{b}_{ab}\\ &=& S_{PF}(\hat{h}) - \frac{M^{2}}{2} \bar{b}^{ab}\bar{b}_{ab}.
 \eea
 
 There are several remarks in order:
 The first one is that  
 the `modified' theory can be written as a usual gravity theory (at least at quadratic level) using the redefinition of the spin 2 field $h \to \hat{h}$. In this version the field $\bar{b}$ 
enters purely algebraically and can be integrated out.
So the modification of gravity that is obtained in pure gravity  is a mere field redefinition.

When one couple the theory to matter fields this field redefinition  becomes relevant 
since the theory obtained  by adding the coupling $h_{\mu \nu} T^{\mu \nu}$ differs from the theory 
obtained by adding the coupling $\hat{h}_{\mu \nu} T^{\mu \nu}$. 
If the coupling of matter is obtained via $h$ as then the theory is indeed modified but what is modified is not really gravity by itself but the way matter couple to gravity.

 The second remark  concerns the fact that
 if matter coupling is via $h$ we can integrate out  the $b$ field since we are working in the quadratic approximation.
 This integration modify the gravity kinetic term, we obtain a non local effective action which differs from the Pauli-Fierz form. 
 The transverse traceless mode 
 acquire a kinetic term of the form
 \be
 h^{TT}_{\mu \nu}  \frac{\Box}{1+\frac{\Box}{M^{2}}} {h^{TT}}^{\mu \nu}.
\ee
The modification due to the presence of a non trivial potential term is 
important only for ultraviolet modes $p^{2}\> \alpha$. The propagator is given by 
$\frac1{p^{2}} + \frac1{M^{2}}$ and the modification is rendering the ultraviolet problem even worse.
We can see that at this quadratic level the modification amounts to add on top of the usual newtonian potential a contact term $V(x) \sim \frac1\alpha \delta^{4}(x)$.

The third remark concerns the fact that the cancellation of the kinetic term for the $b$ field is in agreement with the claim made in \cite{BengH, KirillH} that the modified gravity theory contains as many degrees of freedom as usual gravity. 
the non metric fields $b$ are indeed non propagating at least at the quadratic level. This can be understood from the fact that the action (\ref{SBF2}) without the Potential term is just SU(2)
$BF$ theory. This theory possess an extra topological symmetry labelled by an $SU(2)$ valued one form field $\phi^{j}_{\mu}$, $\delta B^{i}_{\mu \nu} = D_{[\mu} \Phi_{\nu]}^{i}$.
Four components of $\phi^{j}_{\mu}$ can be identify with diffeomorphisms, this left us with 
$3\times 4 -4=8$ gauge symmetries. This  is a priori enough to get rid of the $8$ components of 
$b_{ab}$. 

\section{Generalisation}\label{generalisation}

Finally, it is interesting to note here that the mechanism at work, which allows to deform gravity while keeping the same number of degree of freedom can in fact be generalised. What we need is to introduce a new field which couple to the metric in a covariant way 
while substracting a kinetic term for this new field which insure that this  field is in fact not propagating, and leads to  purely algebraic 
equations of motion for this field.
Such a mechanism in the context where the additional scalar field is a scalar has already been proposed and studied in \cite{Carroll}.
Here we illustrate this general procedure in the case the additional field is a spin 2 field $\pi_{\mu\nu}$.
Lets consider the following Lagrangian 
\be
S_{1}(g,\pi) = \int \sqrt{g} G_{\mu\nu}( g^{\mu\nu} + \pi^{\mu\nu}) \
\ee
where $G_{\mu\nu}=R_{\mu\nu}-\frac{g_{\mu\nu}}2 R$ is the Einstein tensor.
Such lagrangian where studied in \cite{Ovrut} in the context of massive gravity and describe a coupling of a spin two massless field to 
gravity.
The variation with respect to $g$ is given by
\be
\delta g_{\mu\nu}\left(G^{\mu\nu} -\frac12\left( \Box\pi^{\mu\nu} + g^{\mu\nu} \nabla^{\alpha}\nabla^{\beta} \pi^{\alpha \beta} - \nabla_{\alpha}\nabla^{\mu} \pi^{\alpha \nu} 
- \nabla_{\alpha}\nabla^{\nu} \pi^{\alpha \mu} + g^{\mu\nu} \Box \pi - \nabla^{\mu}\nabla^{\nu} \pi + R\pi^{\mu\nu} \right) \right)\nonumber
\ee
We need to choose a kinetic term for the spin two field, and we take a covariant version of the Pauli-Fierz Lagrangian
\be
S_{2}=\int \sqrt{g}\frac14 \left(-\nabla_{\mu}\pi \nabla^{\mu}\pi + \nabla_{\mu}\pi_{\nu\alpha}\nabla^{\mu}\pi^{\nu\alpha}
+ 2 \nabla_{\alpha}\pi_{\nu}^{\alpha}\nabla^{\nu}\pi -2 \nabla_{\mu}\pi_{\nu\alpha}\nabla^{\alpha}\pi^{\mu\nu}\right)
\ee
If one add to this action  a potential term 
\be
S_{3}=\int\sqrt{g} V(\pi_{\mu\nu} \pi^{\mu \nu}, \pi^{2})
\ee
which has a minimum around $\pi_{\mu\nu}=0$.

Then one can check that if one look around the quadratic fluctuation around a solution of Einstein equation
 $g_{\mu\nu}= \bar{g}_{\mu\nu} + h_{\mu\nu}$ the action possesses only one propagating graviton.
 In fact, we can  redefine the fluctuation field $ \hat{h}_{\mu\nu} \equiv h_{\mu\nu} +\pi_{\mu\nu}$ such that the total action is 
 the Pauli-Fierz action for $\hat{h}$ plus a purely algebraic action for $\pi_{\mu\nu}$. Therefore, as in the previous case, the 
 field $\pi$ is not propagating, and the theory possesses only two propagating degree of freedom, despite the fact that 
 after integration of the field $\pi$ this is a non local action for the metric $g$.
 This property comes from the fact that we have precisely tuned the parameter in front of the
 Pauli-Fierz action. 
 In the previous case this tuning  was 
protected by the extra topological BF symmetry.  

 \begin{acknowledgments}
We thank Kirill Krasnov for discussions on this subject.
Research at Perimeter Institute is supported by the Government of Canada through Industry Canada and by the Province of Ontario through the Ministry of Research \& Innovation. 
\end{acknowledgments}

\end{document}